\documentclass[aps,twocolumn,prl]{revtex4}
\usepackage{graphicx}
\usepackage{bm}
\usepackage{hyperref}
\usepackage{color}
\usepackage{amsfonts}

\begin{document}
\newcommand{\St}{St} 
\newcommand{\Dc}{{D_2}}
\newcommand{\TODO}[1]{\textcolor{red}{#1}}
\def\NOTE#1{{\textcolor{red}{\bf [#1]}}}   % note
\def\DEL#1{{\textcolor{green}{#1}}}        % suggested deletions
\def\ADD#1{{\textcolor{blue}{#1}}}         % addition
\def\AD#1{{\textcolor{magenta}{#1}}}       % change
\def\ARED#1{{\textcolor{red}{#1}}}        % question

\newcommand{\ocaaddress}{Laboratoire Lagrange, UMR7293,
  Universit\'e de Nice Sophia-Antipolis, CNRS, Observatoire de la
  C\^ote d'Azur, CS\,34229, 06304 Nice Cedex 4, France}
 
%%%
\title{Clustering, fronts, and heat transfer in turbulent suspensions of heavy particles}
\author{J\'er\'emie Bec}
\affiliation{\ocaaddress}
\author{Holger Homann}
\affiliation{\ocaaddress}
\author{Giorgio Krstulovic}
\affiliation{\ocaaddress}
\begin{abstract}
  Heavy inertial particles transported by a turbulent flow are shown
  to concentrate in the regions where an advected passive scalar, such
  as temperature, displays very strong front-like discontinuities.
  This novel effect is responsible for extremely high levels of
  fluctuations for the passive field sampled by the particles that
  impacts the heat fluxes exchanged between the particles and the
  surrounding fluid.  Instantaneous and averaged heat fluxes are shown
  to follow strongly intermittent statistics and anomalous scaling
  laws.
\end{abstract}
\maketitle
%%%
\noindent 
It is today established that small heavy particles suspended in a
turbulent flow distribute in a strongly non-homogeneous
manner. Quantifying this phenomenon found important applications in
the study of cloud precipitation \cite{falkovich2002acceleration} and
planet formation \cite{johansen2007rapid}. Heavy particles are ejected
by inertial centrifugal forces from vortices and form
\emph{preferential concentrations}. Consequently, they sample the
underlying flow in a very non-uniform manner.  This can have important
consequences when the particles interact with a transported scalar
field, such as the temperature, a supersaturated vapor field or a
pollutant concentration. Such passively transported fields develop
non-trivial geometrical and statistical properties, displaying
anomalous scaling laws~\cite{shraiman2000scalar}.  Turbulence indeed
creates \emph{fronts} across which the scalar strongly varies on very
small lengthscales~\cite{celani2000universality}. Such
quasi-discontinuities appear at the boundaries between the different
circulation zones of the flow and concentrate diffusion. Mixing is
weakened in preferential concentrations and enhanced in fronts. While
these two kinds of inhomogeneities result from turbulent eddies, very
little is known on how they relate and alter the mass and heat
transfer properties of the dispersed phase.

To address such issues we consider a passive scalar field $\theta$
evolving according to the advection-diffusion equation
\begin{equation}
  \partial_t \theta + \bm u \cdot \nabla\theta = \kappa \nabla^2\theta
  + \varphi,
  \label{eq:scalar}
\end{equation}
where $\bm u(\bm x,t)$ is a stationary homogeneous and isotropic
turbulent velocity field solving the three-dimensional incompressible
Navier--Stokes equation, $\kappa$ is the diffusivity and $\varphi(\bm
x,t)$ is a large-scale force. In many physical situations, for
instance in clouds or in convection experiments, there is an imposed
mean scalar gradient $\bm G$. This gradient, that can be taken into
account by setting $\varphi=-\bm G \cdot \bm u$, breaks the isotropy
of the system. However, it is known that the scaling properties of a
passive scalar are universal and do not depend on the large-scale
forcing \cite{celani2001fronts}. Therefore in this Letter, unless
explicitly mentioned, we use a large-scale white-noise in time forcing
in order to preserve isotropy.
At the same time we solve (\ref{eq:scalar}), we consider heavy
inertial (point) particles which experience a viscous drag with the
velocity field $\bm u$. Their individual trajectories are given by
\begin{equation}
  % \frac{\mathrm{d}\bm X_\mathrm{p}}{\mathrm{d}t} = \bm V_\mathrm{p},
  % \quad \frac{\mathrm{d}\bm V_\mathrm{p}}{\mathrm{d}t} =
  % -\frac{1}{\tau_\mathrm{p}}\left[\bm V_\mathrm{p}-\bm u(\bm X_\mathrm{p},t)
  % \right],
  \dot{\bm X}_\mathrm{p} = \bm V_\mathrm{p},
  \quad \dot{\bm V}_\mathrm{p} =
  -\frac{1}{\tau_\mathrm{p}}\left[\bm V_\mathrm{p}-\bm u(\bm X_\mathrm{p},t)
  \right],
  \label{eq:particle}
\end{equation}
where dots designate time derivatives. The relaxation time reads
$\tau_\mathrm{p} = 2\rho_\mathrm{p}a^2/(9\rho_\mathrm{f}\nu)$,
$\rho_\mathrm{p}$ and $\rho_\mathrm{f}$ being the particle and fluid
mass density respectively, $a$ the particle radius and $\nu$ the fluid
kinematic viscosity. Particle inertia is measured in terms of the
\emph{Stokes number} $\St = \tau_\mathrm{p}/\tau_\eta$, where
$\tau_\eta$ designates the turnover time associated to the Kolmogorov
dissipative scale $\eta$ (the smallest active scale of the turbulent
flow).  The case $\St=0$ corresponds to tracers (inertia-less
particles), whose dynamics is $\dot{\bm X}_\mathrm{p} = \bm u(\bm
X_\mathrm{p},t)$.

We make use of direct numerical simulations of the incompressible
Navier--Stokes equations with a large-scale forcing. The velocity
field $\bm u$ and the advected passive scalar $\theta$ are obtained by
the (standard) pseudo-spectral code
\emph{LaTu}~\cite{homann2007impact} using $512^3$ and $1024^3$
grid-points (corresponding to Taylor-scale Reynolds numbers $R_\lambda
= 180$ and $315$). We consider a scalar field of Schmidt number one
($\kappa=\nu$). The values of the different fields at the particle
positions are obtained by tri-cubic interpolation.  The Lagrangian
trajectories of millions of particles with different values of the
Stokes number $\St$ are integrated simultaneously. After a transient,
the full system reaches a statistical stationary state and all results
of this Letter refer to this regime

\begin{figure}[h]
  \includegraphics[width=.7\columnwidth]{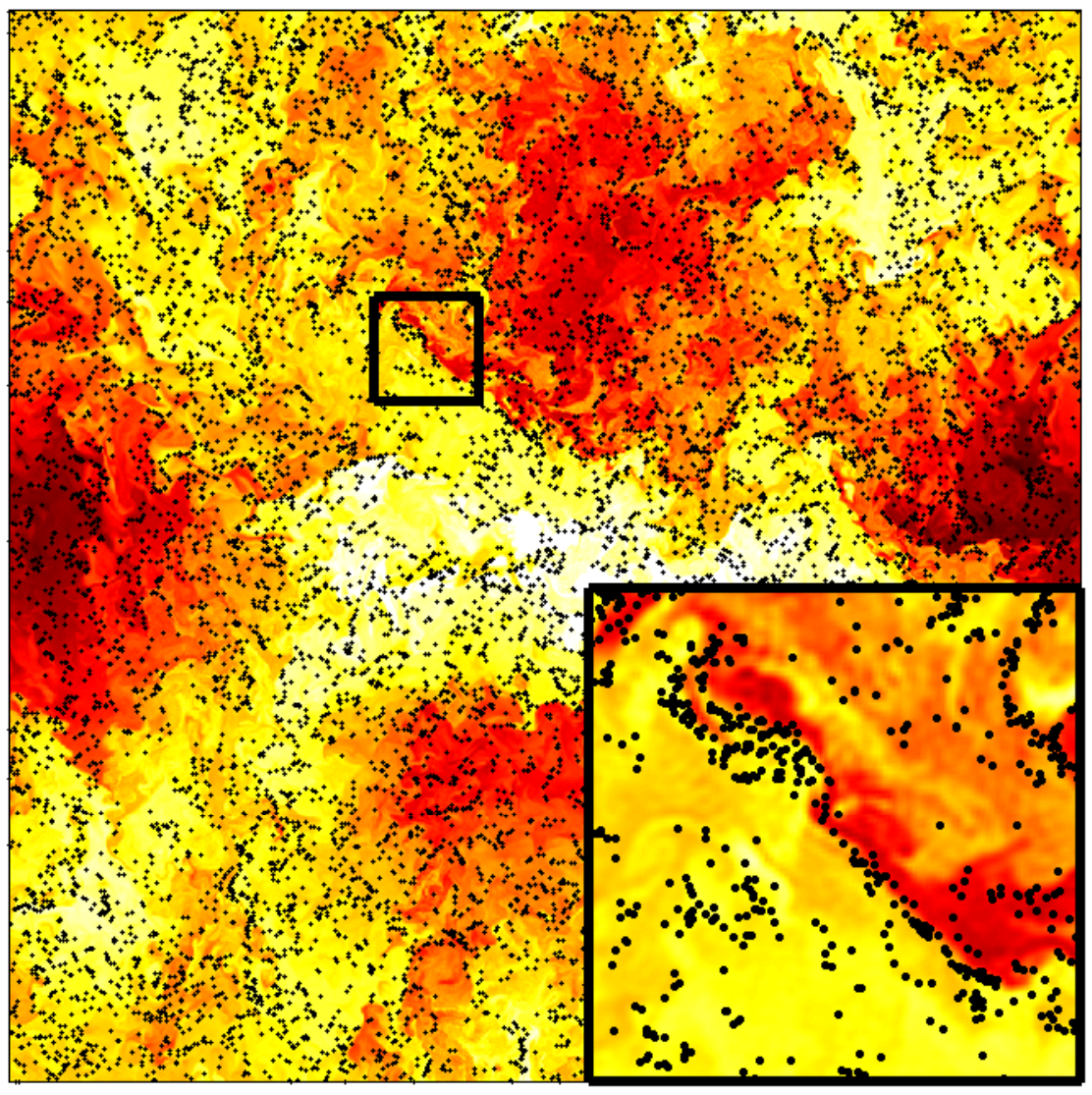}
  \includegraphics[width=.7\columnwidth]{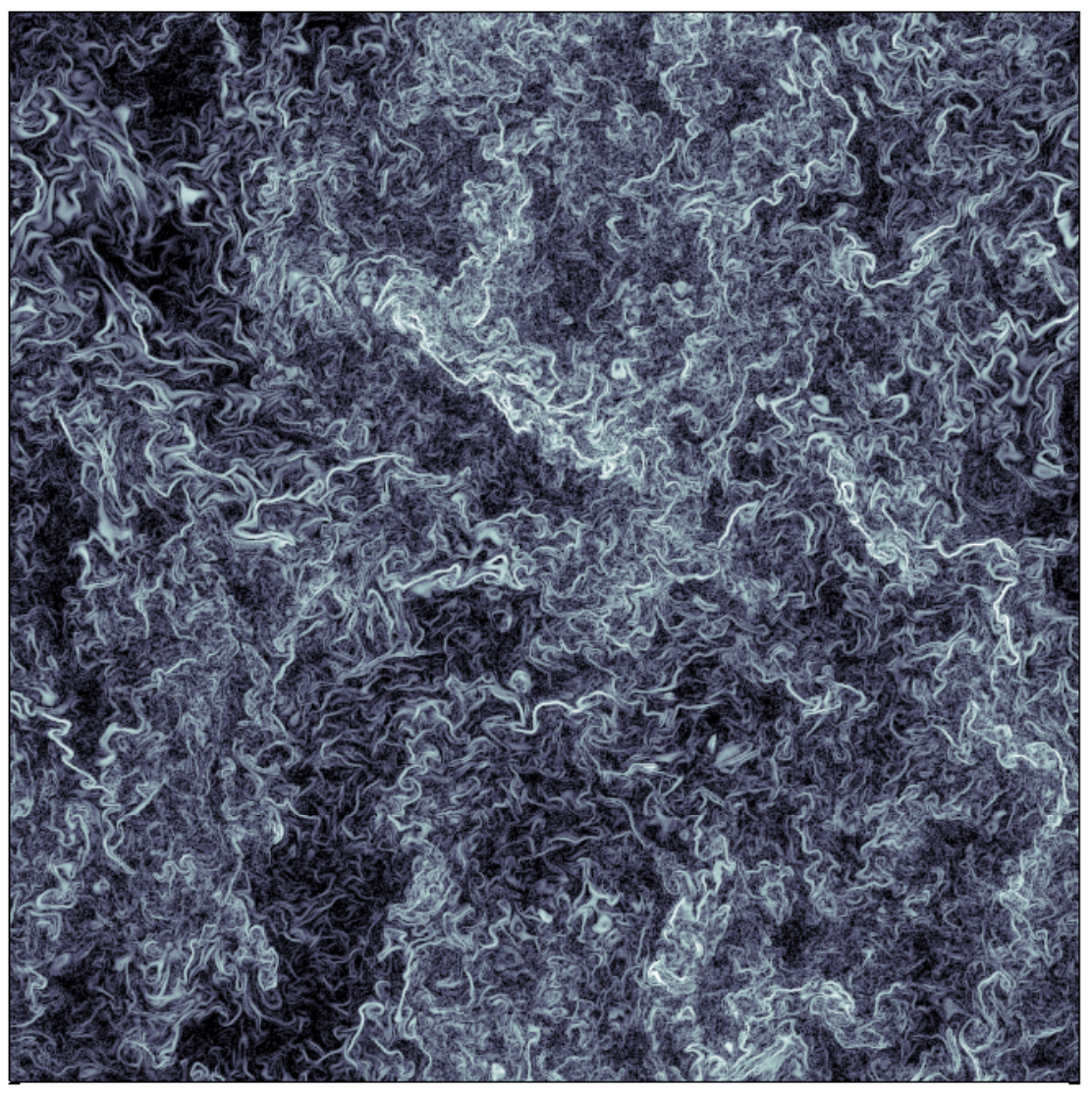} 
  \caption{(color online) \emph{Top}: Snapshot of the scalar field $\theta$ (from
    white to dark red), together with the positions of $\St=1$
    particles (black dots) in a thin slice of width $\simeq\eta$ at
    $R_\lambda \approx 315$. The lower-right inset shows a zoom of the
    black box. \emph{Bottom}: Corresponding snapshot of the scalar
    dissipation rate $\epsilon_\theta$ (from black to white).}
  \label{fig:snapshot}
\end{figure}
\begin{figure}[h]
  \includegraphics[width=.3\columnwidth]{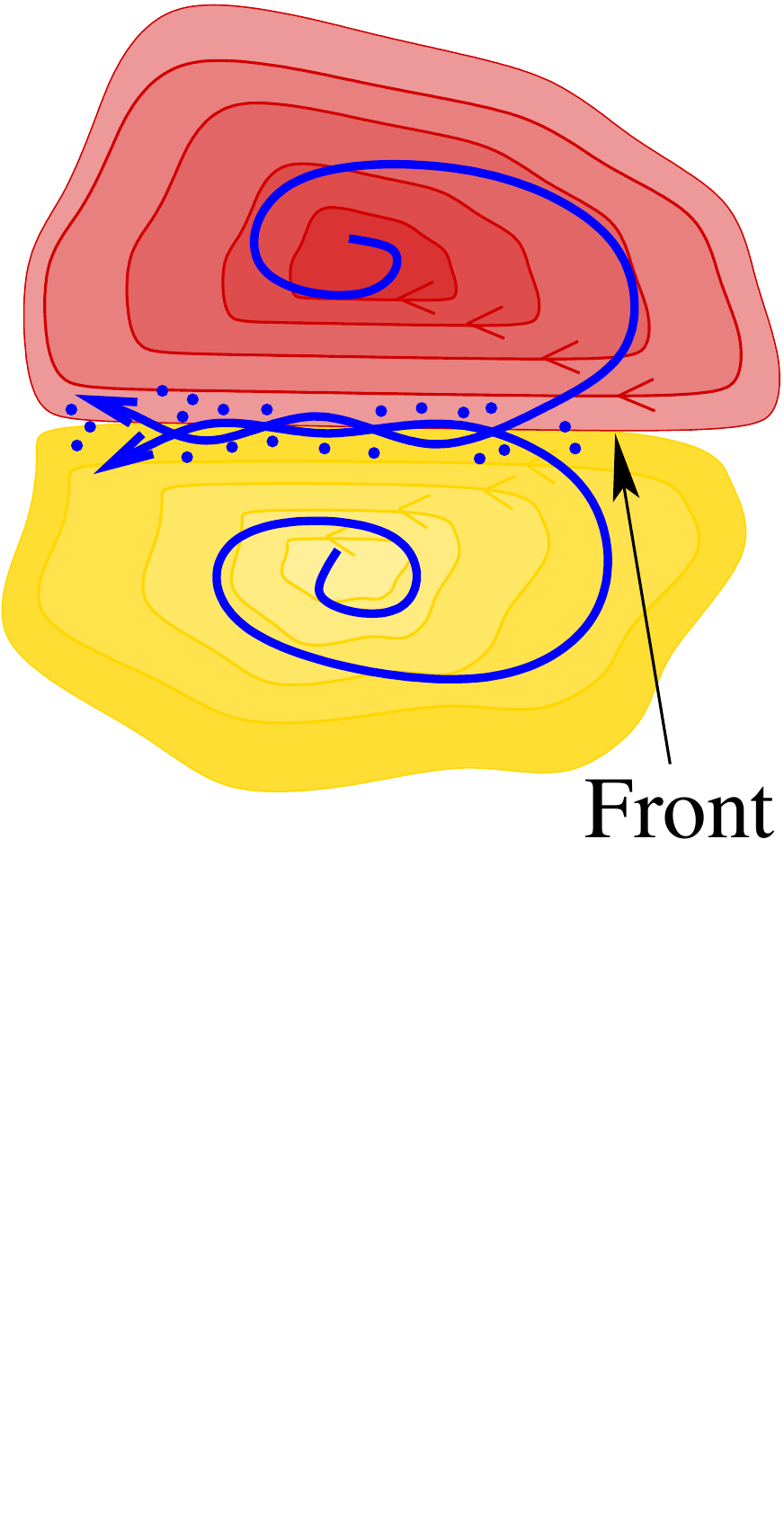}
  \hfill
  \includegraphics[width=.67\columnwidth]{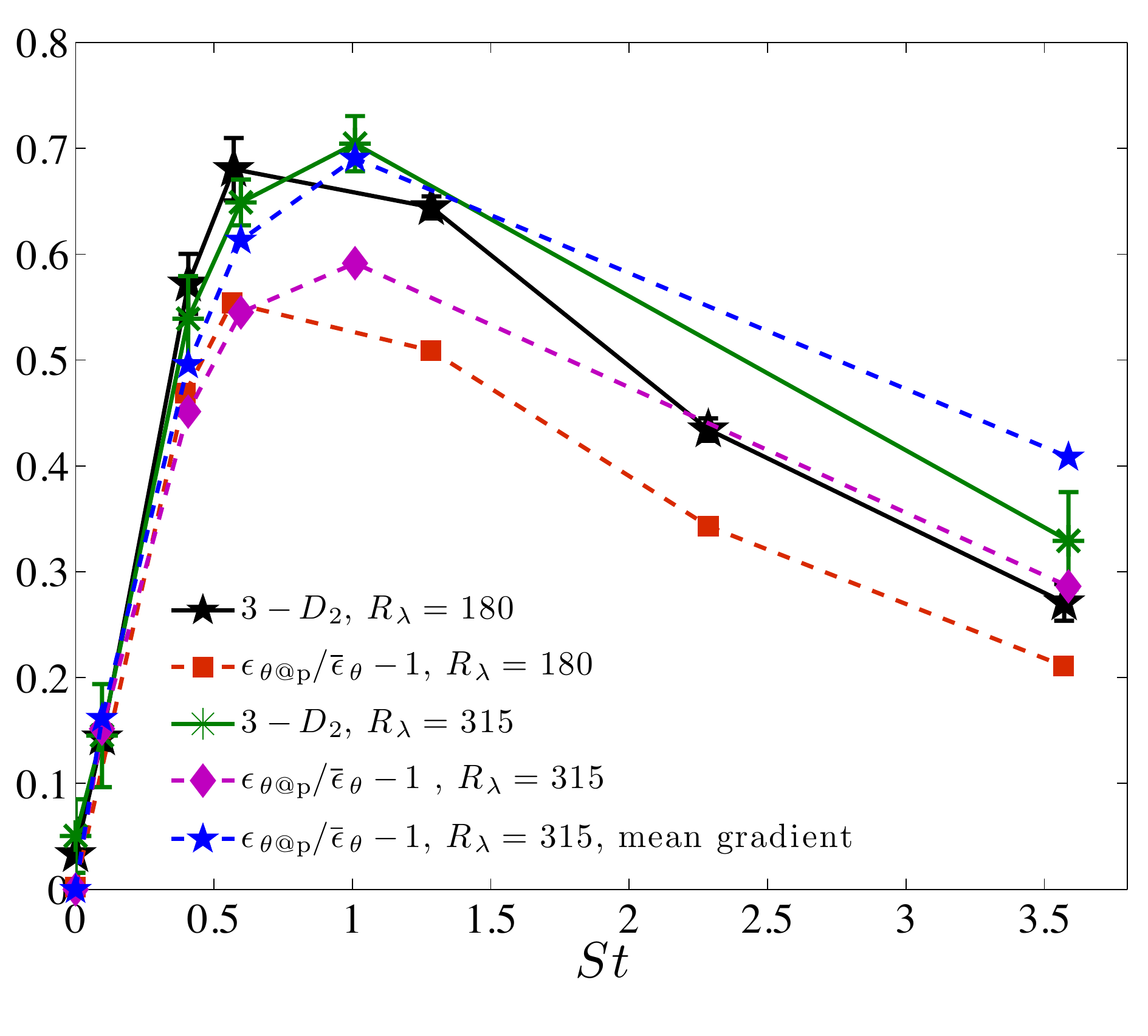}
  \vspace{-10pt}
  \caption{(color online) \emph{Left:} Sketch of the mechanism leading to the
    concentration of particles in the fronts of the scalar field. The
    red and yellow areas outline two different Lagrangian coherent
    structures and the blue lines show typical particle
    trajectories. \emph{Right:} Complementary correlation dimension
    $3-\Dc$ and relative enhancement of scalar dissipation rate at
    particle positions $\epsilon_{\theta @
      p}/\overline{\epsilon_{\theta}}-1$ as a function of the Stokes
    number.}
  \label{fig:clustering}
\end{figure}

Figure~\ref{fig:snapshot} (top) shows a snapshot of the passive scalar
field together with particle positions in a thin slice of the
three-dimensional domain. The scalar field is characterized by the
presence of large-scale regions where it varies smoothly separated by
sharp fronts where it varies abruptly. Particles form clusters that
display a strong correlation with these fronts as emphasized in the
inset.  The regions where the scalar is almost constant are the
so-called \emph{Lagrangian coherent structures} of the flow
\cite{mathur2007uncovering}. They relate to zones where the mixing is
ineffective and thus consist of fluid elements sharing a common
history of the scalar forcing along their paths. Fronts appear at the
border between such closed dynamical regions.  The inertial
centrifugal forces acting on heavy particles are responsible for their
ejection from these regions and their concentration at the edges,
whence the correlations appearing between particle clusters and the
fronts of the scalar field. This mechanism is sketched in
Fig.~\ref{fig:clustering} left.

This effect can also be understood by local arguments. The fronts
correspond to locations where the scalar dissipation rate
$\epsilon_\theta=(\kappa/2) |\nabla \theta|^2$ is very strong.  These
violent spatial fluctuations are clearly appreciated in
Fig.~\ref{fig:snapshot} (bottom) where $\epsilon_\theta$ is displayed
for the same snapshot.  It is easily seen from (\ref{eq:scalar}) that
$\epsilon_\theta$ is stretched by the velocity gradients. Namely, when
neglecting the diffusive and forcing terms, the scalar dissipation
along tracer trajectories obeys $ \dot{\epsilon}_\theta =
-\kappa\,(\nabla\theta)^\mathsf{T} \mathbb{S} (\bm X_\mathrm{p},t)
\nabla\theta$, where $\mathbb{S}$ is the symmetric part of the fluid
velocity gradient tensor $\nabla \bm u$.  This results in an
enhancement of dissipation in the regions where the fluid flow has a
large contraction rate. At the same time, large values of the shear
rate $\mathbb{S}$ enhance the concentration of particles.  As shown in
\cite{maxey1987gravitational}, particles with small inertia ($\St\ll
1$) can be approximated as the tracers of a synthetic compressible
velocity field, namely $\dot{\bm X}_\mathrm{p} \approx \bm v(\bm
X_\mathrm{p},t)$ with $\bm v = \bm u- \tau_\mathrm{p} (\partial_t\bm u
+ \bm u\cdot\nabla\bm u)$. For incompressible fluid flows, the
divergence of the velocity field $\bm v$ reads $\nabla\cdot\bm v =
-\tau_\mathrm{p} ( \mathrm{tr}\,\mathbb{S}^2-|\bm \omega|^2/2)$, where
$\bm\omega = \nabla\times\bm u$ is the vorticity. Particles
concentrate in high-strain low-vorticity regions, explaining their
correlation with the high values of $\epsilon_\theta$ and the location
of the fronts.

A quantitative measurement of particle clustering is given by the
correlation dimension $\Dc$, which is estimated by finding the
small-scale algebraic behavior of $P_2(r)\sim r^\Dc$, the probability
to find two particles at a distance less than $r$. The dependence of
the co-dimension $3-\Dc$ on the Stokes number is shown in
Fig.\ref{fig:clustering}.  A non-monotonic behavior with a maximum of
clustering at $\St\sim 1$ is observed as in \cite{bec2007heavy}. As
particles cluster in the fronts, the average scalar dissipation at the
particles position $\epsilon_{\theta @ p}$ is expected to be sensitive
to the Stokes number. This is apparent in Fig.\ref{fig:clustering}
where the relative enhancement of the scalar dissipation rate
$\epsilon_{\theta @ p}/\overline{\epsilon_{\theta}}-1$ is also plotted
($\overline{\epsilon_{\theta}}$ designates here the mean scalar
dissipation). The dissipation along particle trajectories can be 70\%
larger than the average for values of the Stokes number for which
preferential concentration is the strongest. Note that this behavior
is independent of the way the scalar is forced as it is also observed
when an average gradient is imposed.

\begin{figure}[h]
  \includegraphics[width=\columnwidth]{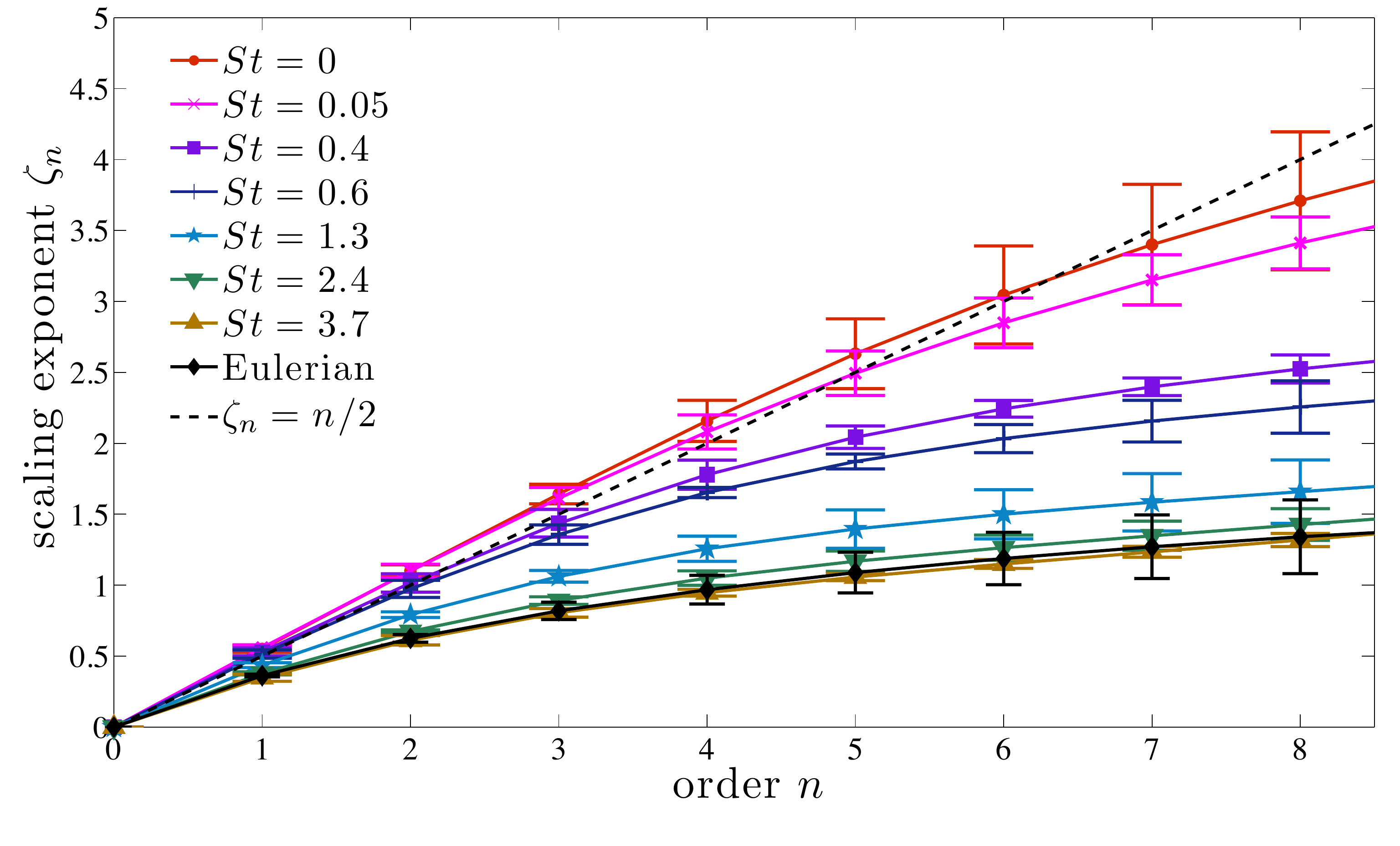}
  \vspace{-10pt}
  \caption{(color online) Anomalous exponents $\zeta_n$ of the Lagrangian increments
    $\langle|\delta_\tau\theta|^n\rangle \sim \tau^{\zeta_n}$ for
    different Stokes numbers. The anomalous exponents $\zeta_n^{\rm E}$
    of the Eulerian increments $\langle|\Delta_\ell
    \theta|^n\rangle\sim \ell^{\zeta_n^{\rm E}}$ are also displayed.}
  \label{fig:exponents}
\end{figure}

We now turn to study the statistics of the passive scalar along
particles trajectories. It is well known that it presents large
fluctuations leading to an anomalous scaling of the Eulerian structure
functions \cite{shraiman2000scalar,celani2000universality}. Here we
focus on the Lagrangian increments of the scalar field
$\delta_\tau\theta = \theta(\bm X_\mathrm{p}(t+\tau),t+\tau)-
\theta(\bm X_\mathrm{p}(t),t)$ that strongly depend on the particles
inertia. For $\St=0$, particles are simple tracers and mainly remain
inside the Lagrangian coherent structures. The variations of
$\theta(\bm X_\mathrm{p}(t))$ are both diffusive due to the forcing
$\varphi$ and relate to the formation, deformation and destruction of
fronts. When $\St>0$, inertia allows particles to cross the fronts and
thus to sample larger fluctuations of the scalar field. When $\St\to
\infty$, the particles decouple from the flow and move almost
ballistically. A frozen Taylor hypothesis leads then to predict that
$\delta_\tau\theta$ is given by the Eulerian increments $\Delta_\ell
\theta =\theta ({\bf x+\ell},t)-\theta ({\bf x},t)$ with $\ell\simeq
\tau v_\mathrm p$, where $v_\mathrm p$ is the typical particle
velocity.
This is manifest when looking at the moments of the Lagrangian
increments that are expected to scale as
$\langle|\delta_\tau\theta|^n\rangle \sim \tau^{\zeta_n}$ for
$\tau_\eta\ll \tau\ll\tau_L$, with $\tau_L$ the large-eddy turnover
time of the turbulent flow. Similarly, the Eulerian increments scales
as $\langle|\Delta_\ell \theta|^n\rangle\sim \ell^{\zeta_n^{\rm E}}$
inside the inertial range ($\eta\ll\ell\ll L$ with $L$ the largest
scale of the system).
Figure~\ref{fig:exponents} shows the scaling exponents $\zeta_n$ as a
function of their order $n$ for various values of the particle Stokes
number $\St$. For tracers ($\St=0$) the results are very close to the
normal scaling $\zeta_n = n/2$, indicating that anomalous corrections,
if any, are very weak and quantifying them precisely would require a
major augmentation of the statistics. When increasing $\St$ the
exponents $\zeta_n$ go from a tracer behavior to those obtained from
Eulerian statistics $\zeta_n=\zeta_n^{\rm E}$, showing the enhancement
of Lagrangian scalar intermittency due to particle inertia.

These findings have important consequences on possible heat exchanges
between particles and a carrier fluid. Indeed, let us assume that the
transported scalar field is the fluid temperature. For particles much
smaller than the scales at which the fluid temperature varies, the
heat flux at the particle surface is proportional to the difference
between the particle temperature $\theta_p$ and that of the
environment \cite{wetchagarun2010dispersion}, so that
\begin{equation}
  \dot{\theta}_\mathrm{p} =
  -\frac{1}{\tau_\theta}\left[\theta_\mathrm{p}-\theta(\bm
    X_\mathrm{p},t)\right],
  \label{eq:temp_part}
\end{equation}
where $\tau_\theta = c_\mathrm{p}a^2/(3c_\mathrm{f}\kappa)$, with
$c_\mathrm{p}$ and $c_\mathrm{f}$ the volumetric heat capacities of
the particles and the fluid respectively. The particles then have a
thermal inertia that is measured in terms of the \emph{thermal Stokes
  number} $\St_\theta = \tau_\theta/\tau_\eta$.

\begin{figure}[h]
  \includegraphics[width=1\columnwidth]{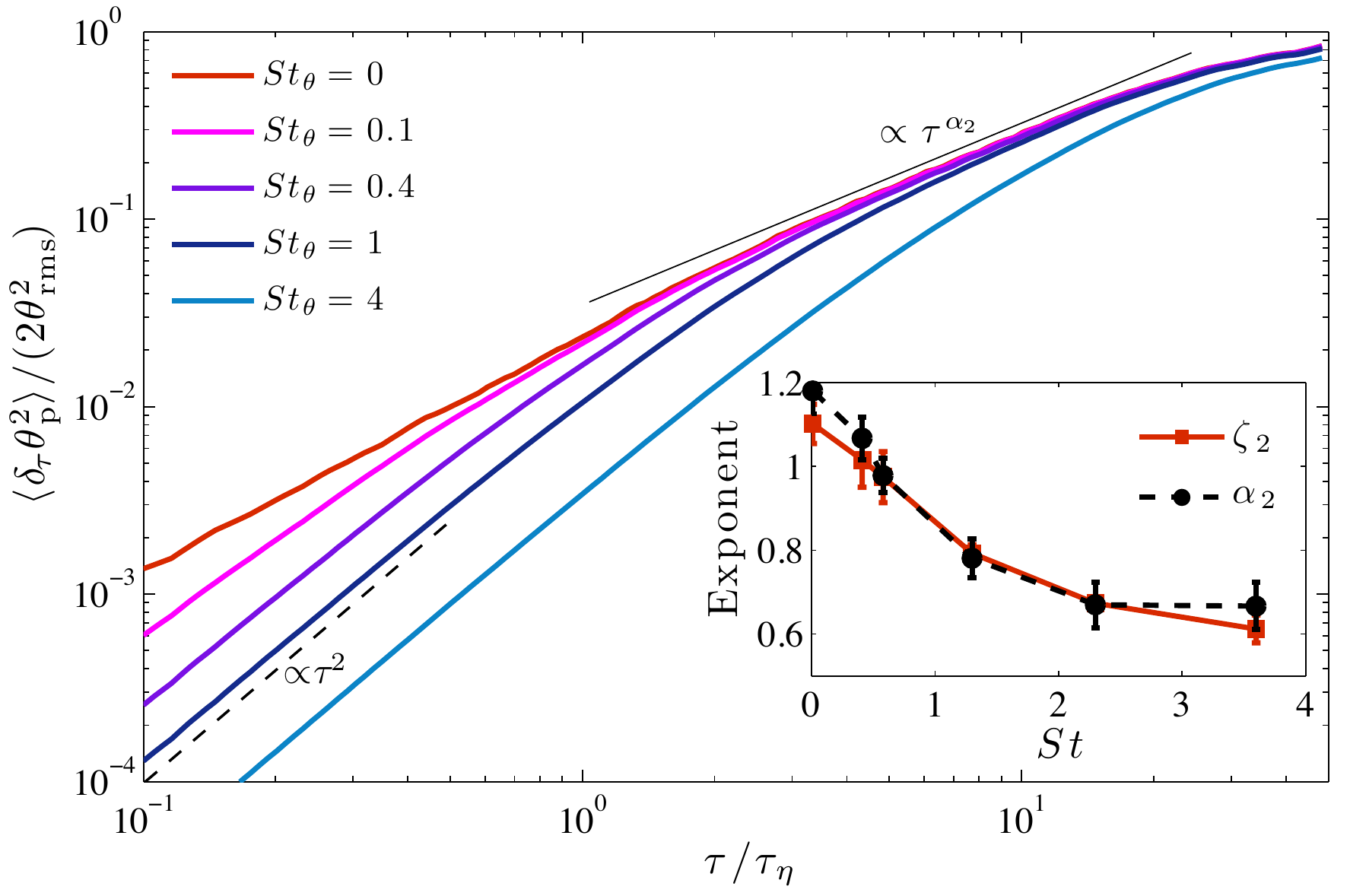}
  \vspace{-10pt}
  \caption{(color online) Time evolution of $\langle \delta_\tau\theta_\mathrm{p}^2
    \rangle$ for $\St = 0.6$ and five different values of the thermal
    Stokes number $\St_\theta$. The inset displays the exponent
    $\alpha_2$ of the inner particle temperature scaling
    $\delta_\tau\theta_p$ for time lags $\tau\gg\tau_\theta$ together
    with the anomalous exponent $\zeta_2$ of the fluid temperature for
    different Stokes numbers.}
  \label{fig:evolDT}
\end{figure}

Heat exchanges between the particles and the fluid are entailed in the
dependence of the particle temperature increment
$\delta_\tau\theta_\mathrm{p} = \theta_\mathrm{p}(t+\tau)-
\theta_\mathrm{p}(t)$ upon the time lag $\tau$. As the system is in a
statistical steady state, the increments are independent of
$t$. Different regimes occur. For small time lags $\tau \ll
\tau_\theta$, the heat flux remains almost constant and
$\theta_\mathrm{p}(t+\tau) \simeq \theta_\mathrm{p}(t) + \tau\,
\dot{\theta}_\mathrm{p}(t)$, so that $\langle
\delta_\tau\theta_\mathrm{p}^n \rangle \simeq \tau^n \langle
\dot{\theta}_\mathrm{p}^n\rangle$. This regime is observed in
Fig.~\ref{fig:evolDT}, which represents the evolution of $\langle
\delta_\tau\theta_\mathrm{p}^2 \rangle$ for $\St=0.6$ and various
values of the particle thermal inertia. At larger time lags, one
observes that temperature increments follow anomalous scaling laws of
the form $\langle \delta_\tau\theta_\mathrm{p}^n \rangle \simeq
\tau^{\alpha_n}$.  This regime occurs when
$\tau_\theta\ll\tau\ll\tau_L$ that is in the limit when thermal
inertia becomes negligible and particle temperature follows that of
the fluid. We then expect $\langle \delta_\tau\theta_\mathrm{p}^n
\rangle\simeq\langle \delta_\tau\theta^n \rangle$, so that the scaling
laws of particle temperature are given by the anomalous Lagrangian
exponents of the scalar field introduced above, namely $\alpha_n
=\zeta_n$. This is confirmed form the inset of Fig.~\ref{fig:evolDT}
where both $\alpha_2$ and $\zeta_2$ are displayed as a function of the
Stokes number.

\begin{figure}[h]
  \includegraphics[width=\columnwidth]{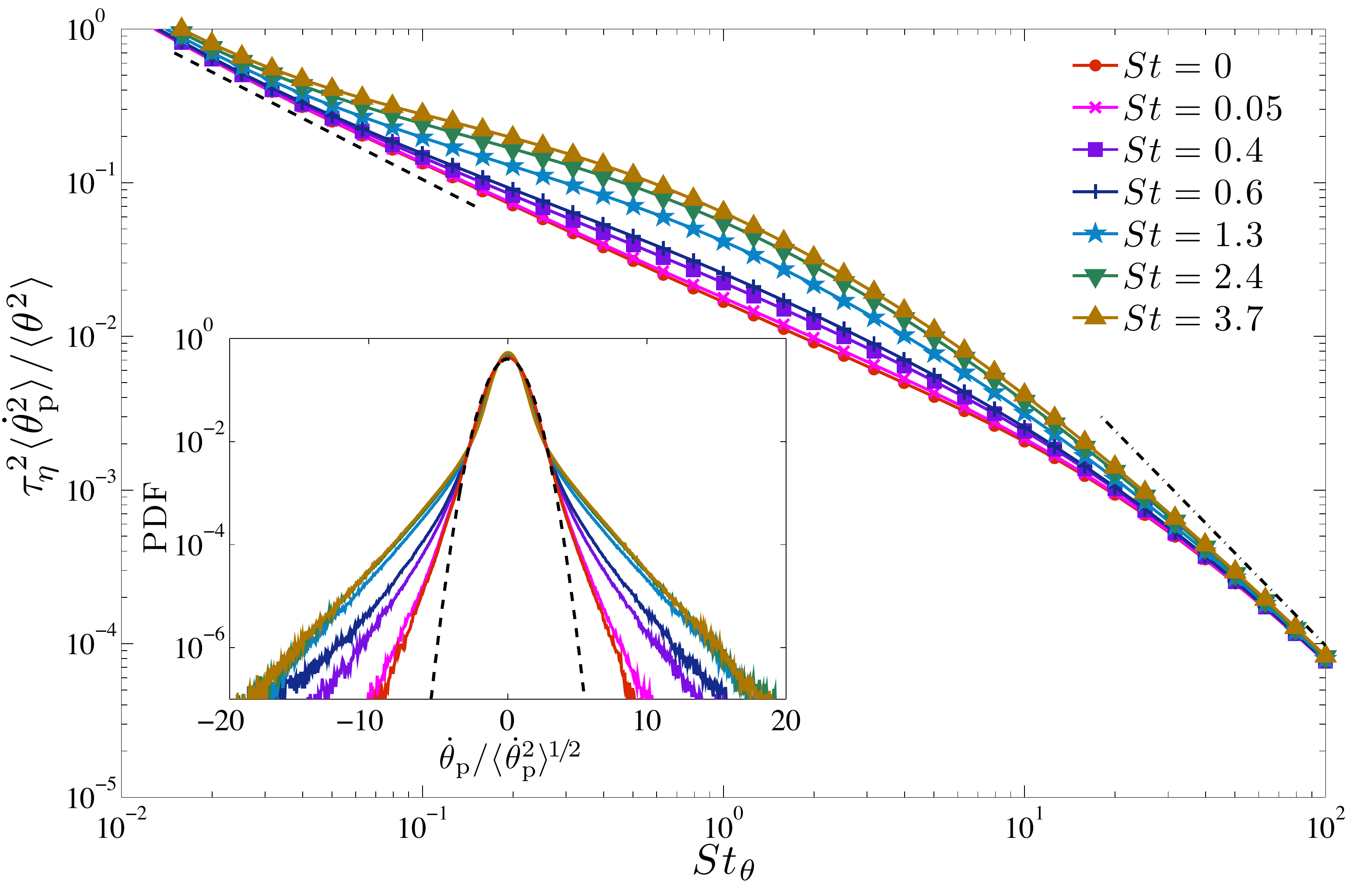}
  \vspace{-10pt}
  \caption{(color online) Variance of the heat flux $\langle
    \dot{\theta}_\mathrm{p}^2\rangle$ as a function of the thermal
    Stokes number $\St_\theta$ and various particle inertia; the two
    lines represents the asymptotics $\langle
    \dot{\theta}_\mathrm{p}^2\rangle \propto \St_\theta^{-1}$ for
    $\St_\theta\ll 1$ (dashed line) and $\langle
    \dot{\theta}_\mathrm{p}^2\rangle \propto \St_\theta^{-2}$ for
    $\St_\theta\gg 1$ (dotted-dashed line). Inset: probability density
    function (PDF) of the heat flux $\dot{\theta}_\mathrm{p}$
    normalized to unit variance for $\St_\theta=1$ and the different
    Stokes numbers; the dashed curve shows a Gaussian distribution.}
  \label{fig:var_flux}
\end{figure}
The instantaneous heat exchanges between the particles and the fluid
are also strongly depending on both the thermal and dynamical
inertia. This is evidenced from Fig.~\ref{fig:var_flux}, which
represents the heat flux variance $\langle\dot{\theta}_\mathrm{p}^2
\rangle$ as a function of the thermal Stokes number $\St_\theta$ and
various values of $\St$.  One clearly observes that when particle
inertia increases, the fluctuations of the heat flux become stronger
with a maximum deviation from tracers when $\St_\theta$ is of the
order of unity. For the largest Stokes number we have investigated
here ($\St=3.7$), one observes at $\St_\theta=1$ a gain of a factor
almost three. The variance of the heat flux can be related to the
Lagrangian fluid temperature increment. One can indeed easily check,
using~(\ref{eq:temp_part}), that statistical stationarity implies
\begin{equation}
  \langle\dot{\theta}_\mathrm{p}^2\rangle = \frac{1}{2\tau_\theta^3}
  \int_0^\infty \langle \delta_\tau\theta^2\rangle
  \,\mathrm{e}^{-\tau/\tau_\theta} \,\mathrm{d}\tau,
  \label{eq:var_flux_dtheta}
\end{equation}
where $\langle \delta_\tau\theta^2\rangle$ designates the second-order
Lagrangian structure function of the fluid temperature along the
particle path and over a time lag $\tau$. When
$\tau_\theta\ll\tau_\eta$ the integral is concentrated on the small
values of $\tau$ where the variations of $\theta$ are dominated by the
$\delta$-correlated in time forcing and thus $\theta$ diffuses and
$\langle \delta_\tau\theta^2\rangle \sim \tau$. A saddle-point
argument then gives $\langle\dot{\theta}_\mathrm{p}^2
\rangle\sim\St_\theta^{-1}$ when $\St_\theta\ll1$, as observed in our
data. Conversely, for extremely large $\tau_\theta$, the integral is
dominated by the large values of $\tau$.  When $\tau\gg\tau_L$, one
expects $\langle \delta_\tau\theta^2\rangle\simeq
2\langle\theta^2\rangle$, so that $\langle\dot{\theta}_\mathrm{p}^2
\rangle \sim \St_\theta^{-2}$ for $\tau_\theta\gg\tau_L$, that is
$\St_\theta\gg R_\lambda/\sqrt{15}$. In between these two asymptotics,
the anomalous scaling $\langle \delta_\tau\theta^2\rangle\sim
\tau^{\zeta_2}$ of the Lagrangian temperature structure function
yields a non-trivial behavior.  Indeed, when $1\ll\St_\theta\ll
R_\lambda/\sqrt{15}$, the main contribution to the integral comes from
$\tau$ in the inertial range. This leads to
$\langle\dot{\theta}_\mathrm{p}^2 \rangle \sim
\St_\theta^{\zeta_2-2}$, giving a behavior that hence depends on the
dynamical Stokes number. The increase of the variance of heat flux as
a function of the particle inertia is thus directly related to the
enhancement of Lagrangian scalar intermittency. This effect is of
course not limited to second-order statistics, as illustrated in the
inset of Fig.~\ref{fig:var_flux}. The probability density functions of
$\dot{\theta}_\mathrm{p}$ normalized to a unit variance strongly
depart from a Gaussian and develop fatter and fatter tails when $\St$
increases. This is again a signature of the intermittency of the
scalar field sampled by inertial particles.

The clustering of particles in the temperature fronts and the
resulting anomalous scaling laws that are found here reveal that a
dispersed phase participates in an active and possibly controlled
manner to the heat transport in a turbulent flow. Depending on the
values of their dynamical and thermal response times, the particles
can either act as thermostats or accelerate the diffusion of
temperature in the fluid. Such properties can be used to ameliorate
and optimize the design of numerous industrial devices ranging from
combustion engines to chemical reactors. Furthermore, in addition to
thermal properties, the mass transfers between the fluid and the
particles are also ruled by the intermittent effects unveiled
here. Our results indicate that droplets in turbulent clouds as well
as dust particles in protoplanetary disks concentrate at the
boundaries between wet and dry, dense and sparse regions. Their
inertia allows them to travel across such high-variability zones and
thus to experience very different growth histories by condensation or
accretion. We expect this effect to be responsible for a critical
broadening of the particle size distribution that is not predicted by
classical mean-field kinetic approaches. This effect could partly
explain the difficulties encountered when estimating the timescales of
both rain and planet formation.

This work was performed using HPC resources from GENCI-TGCC (Grant
2013-2b6815). The research leading to these results has received
funding from the European Research Council under the European
Community's 7th Framework Program (FP7/2007-2013, Grant Agreement
no. 240579) and from the Agence Nationale de la Recherche (Programme
Blanc ANR-12-BS09-011-04).

\bibliographystyle{apsrev4-1}
\bibliography{biblio}

%merlin.mbs apsrev4-1.bst 2010-07-25 4.21a (PWD, AO, DPC) hacked
%Control: key (0)
%Control: author (72) initials jnrlst
%Control: editor formatted (1) identically to author
%Control: production of article title (-1) disabled
%Control: page (0) single
%Control: year (1) truncated
%Control: production of eprint (0) enabled
\begin{thebibliography}{10}%
\makeatletter
\providecommand \@ifxundefined [1]{%
 \@ifx{#1\undefined}
}%
\providecommand \@ifnum [1]{%
 \ifnum #1\expandafter \@firstoftwo
 \else \expandafter \@secondoftwo
 \fi
}%
\providecommand \@ifx [1]{%
 \ifx #1\expandafter \@firstoftwo
 \else \expandafter \@secondoftwo
 \fi
}%
\providecommand \natexlab [1]{#1}%
\providecommand \enquote  [1]{``#1''}%
\providecommand \bibnamefont  [1]{#1}%
\providecommand \bibfnamefont [1]{#1}%
\providecommand \citenamefont [1]{#1}%
\providecommand \href@noop [0]{\@secondoftwo}%
\providecommand \href [0]{\begingroup \@sanitize@url \@href}%
\providecommand \@href[1]{\@@startlink{#1}\@@href}%
\providecommand \@@href[1]{\endgroup#1\@@endlink}%
\providecommand \@sanitize@url [0]{\catcode `\\12\catcode `\$12\catcode
  `\&12\catcode `\#12\catcode `\^12\catcode `\_12\catcode `\%12\relax}%
\providecommand \@@startlink[1]{}%
\providecommand \@@endlink[0]{}%
\providecommand \url  [0]{\begingroup\@sanitize@url \@url }%
\providecommand \@url [1]{\endgroup\@href {#1}{\urlprefix }}%
\providecommand \urlprefix  [0]{URL }%
\providecommand \Eprint [0]{\href }%
\providecommand \doibase [0]{http://dx.doi.org/}%
\providecommand \selectlanguage [0]{\@gobble}%
\providecommand \bibinfo  [0]{\@secondoftwo}%
\providecommand \bibfield  [0]{\@secondoftwo}%
\providecommand \translation [1]{[#1]}%
\providecommand \BibitemOpen [0]{}%
\providecommand \bibitemStop [0]{}%
\providecommand \bibitemNoStop [0]{.\EOS\space}%
\providecommand \EOS [0]{\spacefactor3000\relax}%
\providecommand \BibitemShut  [1]{\csname bibitem#1\endcsname}%
\let\auto@bib@innerbib\@empty
%</preamble>
\bibitem [{\citenamefont {Falkovich}\ \emph {et~al.}(2002)\citenamefont
  {Falkovich}, \citenamefont {Fouxon},\ and\ \citenamefont
  {Stepanov}}]{falkovich2002acceleration}%
  \BibitemOpen
  \bibfield  {author} {\bibinfo {author} {\bibfnamefont {G.}~\bibnamefont
  {Falkovich}}, \bibinfo {author} {\bibfnamefont {A.}~\bibnamefont {Fouxon}}, \
  and\ \bibinfo {author} {\bibfnamefont {M.}~\bibnamefont {Stepanov}},\
  }\href@noop {} {\bibfield  {journal} {\bibinfo  {journal} {Nature}\ }\textbf
  {\bibinfo {volume} {419}},\ \bibinfo {pages} {151} (\bibinfo {year}
  {2002})}\BibitemShut {NoStop}%
\bibitem [{\citenamefont {Johansen}\ \emph {et~al.}(2007)\citenamefont
  {Johansen}, \citenamefont {Oishi}, \citenamefont {Mac~Low}, \citenamefont
  {Klahr}, \citenamefont {Henning},\ and\ \citenamefont
  {Youdin}}]{johansen2007rapid}%
  \BibitemOpen
  \bibfield  {author} {\bibinfo {author} {\bibfnamefont {A.}~\bibnamefont
  {Johansen}}, \bibinfo {author} {\bibfnamefont {J.~S.}\ \bibnamefont {Oishi}},
  \bibinfo {author} {\bibfnamefont {M.-M.}\ \bibnamefont {Mac~Low}}, \bibinfo
  {author} {\bibfnamefont {H.}~\bibnamefont {Klahr}}, \bibinfo {author}
  {\bibfnamefont {T.}~\bibnamefont {Henning}}, \ and\ \bibinfo {author}
  {\bibfnamefont {A.}~\bibnamefont {Youdin}},\ }\href@noop {} {\bibfield
  {journal} {\bibinfo  {journal} {Nature}\ }\textbf {\bibinfo {volume} {448}},\
  \bibinfo {pages} {1022} (\bibinfo {year} {2007})}\BibitemShut {NoStop}%
\bibitem [{\citenamefont {Shraiman}\ and\ \citenamefont
  {Siggia}(2000)}]{shraiman2000scalar}%
  \BibitemOpen
  \bibfield  {author} {\bibinfo {author} {\bibfnamefont {B.~I.}\ \bibnamefont
  {Shraiman}}\ and\ \bibinfo {author} {\bibfnamefont {E.~D.}\ \bibnamefont
  {Siggia}},\ }\href {http://dx.doi.org/10.1038/35015000} {\bibfield  {journal}
  {\bibinfo  {journal} {Nature}\ }\textbf {\bibinfo {volume} {405}},\ \bibinfo
  {pages} {639} (\bibinfo {year} {2000})}\BibitemShut {NoStop}%
\bibitem [{\citenamefont {Celani}\ \emph {et~al.}(2000)\citenamefont {Celani},
  \citenamefont {Lanotte}, \citenamefont {Mazzino},\ and\ \citenamefont
  {Vergassola}}]{celani2000universality}%
  \BibitemOpen
  \bibfield  {author} {\bibinfo {author} {\bibfnamefont {A.}~\bibnamefont
  {Celani}}, \bibinfo {author} {\bibfnamefont {A.}~\bibnamefont {Lanotte}},
  \bibinfo {author} {\bibfnamefont {A.}~\bibnamefont {Mazzino}}, \ and\
  \bibinfo {author} {\bibfnamefont {M.}~\bibnamefont {Vergassola}},\ }\href
  {http://dx.doi.org/10.1103/PhysRevLett.84.2385} {\bibfield  {journal}
  {\bibinfo  {journal} {Phys. Rev. Lett.}\ }\textbf {\bibinfo {volume} {84}},\
  \bibinfo {pages} {2385} (\bibinfo {year} {2000})}\BibitemShut {NoStop}%
\bibitem [{\citenamefont {Celani}\ \emph {et~al.}(2001)\citenamefont {Celani},
  \citenamefont {Lanotte}, \citenamefont {Mazzino},\ and\ \citenamefont
  {Vergassola}}]{celani2001fronts}%
  \BibitemOpen
  \bibfield  {author} {\bibinfo {author} {\bibfnamefont {A.}~\bibnamefont
  {Celani}}, \bibinfo {author} {\bibfnamefont {A.}~\bibnamefont {Lanotte}},
  \bibinfo {author} {\bibfnamefont {A.}~\bibnamefont {Mazzino}}, \ and\
  \bibinfo {author} {\bibfnamefont {M.}~\bibnamefont {Vergassola}},\
  }\href@noop {} {\bibfield  {journal} {\bibinfo  {journal} {Physics of
  Fluids}\ }\textbf {\bibinfo {volume} {13}},\ \bibinfo {pages} {1768}
  (\bibinfo {year} {2001})}\BibitemShut {NoStop}%
\bibitem [{\citenamefont {Homann}\ \emph {et~al.}(2007)\citenamefont {Homann},
  \citenamefont {Dreher},\ and\ \citenamefont {Grauer}}]{homann2007impact}%
  \BibitemOpen
  \bibfield  {author} {\bibinfo {author} {\bibfnamefont {H.}~\bibnamefont
  {Homann}}, \bibinfo {author} {\bibfnamefont {J.}~\bibnamefont {Dreher}}, \
  and\ \bibinfo {author} {\bibfnamefont {R.}~\bibnamefont {Grauer}},\
  }\href@noop {} {\bibfield  {journal} {\bibinfo  {journal} {Comp. Phys.
  Comm.}\ }\textbf {\bibinfo {volume} {177}},\ \bibinfo {pages} {560} (\bibinfo
  {year} {2007})}\BibitemShut {NoStop}%
\bibitem [{\citenamefont {Mathur}\ \emph {et~al.}(2007)\citenamefont {Mathur},
  \citenamefont {Haller}, \citenamefont {Peacock}, \citenamefont
  {Ruppert-Felsot},\ and\ \citenamefont {Swinney}}]{mathur2007uncovering}%
  \BibitemOpen
  \bibfield  {author} {\bibinfo {author} {\bibfnamefont {M.}~\bibnamefont
  {Mathur}}, \bibinfo {author} {\bibfnamefont {G.}~\bibnamefont {Haller}},
  \bibinfo {author} {\bibfnamefont {T.}~\bibnamefont {Peacock}}, \bibinfo
  {author} {\bibfnamefont {J.~E.}\ \bibnamefont {Ruppert-Felsot}}, \ and\
  \bibinfo {author} {\bibfnamefont {H.~L.}\ \bibnamefont {Swinney}},\
  }\href@noop {} {\bibfield  {journal} {\bibinfo  {journal} {Phys. Rev. Lett.}\
  }\textbf {\bibinfo {volume} {98}},\ \bibinfo {pages} {144502} (\bibinfo
  {year} {2007})}\BibitemShut {NoStop}%
\bibitem [{\citenamefont {Maxey}(1987)}]{maxey1987gravitational}%
  \BibitemOpen
  \bibfield  {author} {\bibinfo {author} {\bibfnamefont {M.}~\bibnamefont
  {Maxey}},\ }\href@noop {} {\bibfield  {journal} {\bibinfo  {journal} {J.
  Fluid Mech.}\ }\textbf {\bibinfo {volume} {174}},\ \bibinfo {pages} {441}
  (\bibinfo {year} {1987})}\BibitemShut {NoStop}%
\bibitem [{\citenamefont {Bec}\ \emph {et~al.}(2007)\citenamefont {Bec},
  \citenamefont {Biferale}, \citenamefont {Cencini}, \citenamefont {Lanotte},
  \citenamefont {Musacchio},\ and\ \citenamefont {Toschi}}]{bec2007heavy}%
  \BibitemOpen
  \bibfield  {author} {\bibinfo {author} {\bibfnamefont {J.}~\bibnamefont
  {Bec}}, \bibinfo {author} {\bibfnamefont {L.}~\bibnamefont {Biferale}},
  \bibinfo {author} {\bibfnamefont {M.}~\bibnamefont {Cencini}}, \bibinfo
  {author} {\bibfnamefont {A.}~\bibnamefont {Lanotte}}, \bibinfo {author}
  {\bibfnamefont {S.}~\bibnamefont {Musacchio}}, \ and\ \bibinfo {author}
  {\bibfnamefont {F.}~\bibnamefont {Toschi}},\ }\href@noop {} {\bibfield
  {journal} {\bibinfo  {journal} {Phys. Rev. Lett.}\ }\textbf {\bibinfo
  {volume} {98}},\ \bibinfo {pages} {084502} (\bibinfo {year}
  {2007})}\BibitemShut {NoStop}%
\bibitem [{\citenamefont {Wetchagarun}\ and\ \citenamefont
  {Riley}(2010)}]{wetchagarun2010dispersion}%
  \BibitemOpen
  \bibfield  {author} {\bibinfo {author} {\bibfnamefont {S.}~\bibnamefont
  {Wetchagarun}}\ and\ \bibinfo {author} {\bibfnamefont {J.~J.}\ \bibnamefont
  {Riley}},\ }\href {http://dx.doi.org/10.1063/1.3392772} {\bibfield  {journal}
  {\bibinfo  {journal} {Phys. Fluids}\ }\textbf {\bibinfo {volume} {22}},\
  \bibinfo {pages} {063301} (\bibinfo {year} {2010})}\BibitemShut {NoStop}%
\end{thebibliography}%

\end{document}